\documentclass[12pt]{article}
\usepackage{graphicx}
\usepackage{epsf,psfig,amsmath}
\usepackage{epsfig}
\newcommand{\beq}{\begin{equation}}
\newcommand{\eeq}[1]{\label{#1} \end{equation}}

\newcommand{\eq}[1]{\begin{align} #1 \end{align}}

\def\lleq{\lower0.9ex\hbox{$\buildrel < \over \sim$}}
\def\arrow1{\lower0.9ex\hbox{$\buildrel \longrightarrow \over a \to b$}}

\begin{document}

\sloppy


\vskip 10cm \centerline{\bf Duality, analyticity and $t$
dependence of generalized parton distributions} \vskip 0.3cm
\centerline {L\'aszl\'o Jenkovszky$^\star$}

\vskip 0.1cm

\centerline{\sl Bogolyubov Institute for Theoretical Physics,}
\centerline{\sl National Academy of Sciences of Ukraine, Kiev-143,
03680 UKRAINE}

\vskip 1cm

\begin{abstract}

Basing upon dual analytic models, we present arguments in favor of
the parametrization of generalized parton distributions (GPD) in
the form $\sim(x/g_0)^{(1-x)\tilde\alpha(t)},$ where
 $\tilde\alpha(t)=\alpha(t)-\alpha(0)$ is the nonlinear part of
 the Regge trajectory and $g_0$ is a parameter, $g_0>1.$ For linear trajectories
 it reduces to earlier proposals. We compare the calculated moments of
 these GPD with the experimental data on form factors and find that the effects from
the nonlinearity of Regge trajectories are large. By
Fourier-transforming the obtained GPD, we access the spatial
distribution of protons in the transverse plane. The relation
between dual amplitudes with Mandelstam analyticity and composite
models in the infinite momentum frame is discussed, the
integration variable in dual models being associated with the
quark longitudinal momentum fraction $x$ in the nucleon.
\end{abstract}

\vskip 10cm


$
\begin{array}{ll}
^{\star}\mbox{{\it e-mail address:}} & \mbox{jenk@bitp.kiev.ua}
\end{array}
$


\newpage

\section{Introduction} \label{s1}

Generalized parton distributions (GPD) \cite{Mueller, Rad, Ji}
combine our knowledge about the one-dimensional parton
distribution in the longitudinal momentum with the
impact-parameter, or transverse distribution of matter in a hadron
or nucleus. It is an ambitious program to access the spatial
distribution of partons in the transverse plane and thus to
provide a 3-dimensional picture of the nucleon (nucleus)
\cite{Burk1, Burk2, Diehl, Pire, BJY}. This program involves
various approaches, including perturbative QCD, Regge poles,
lattice calculations etc. (see Ref. \cite{review} for reviews).
The main problem is that, while the partonic subprocess can be
calculated perturbatively, the calculation of GPDs require
non-perturbative methods. GPDs enter in hard exclusive processes,
such as deeply virtual Compton scattering (DVCS); however, they
cannot be measured directly but instead appear in convolution
integrals, that cannot be easily converted. Hence the strategy is
to guess the GPD, based on various theoretical constraints, and
then compare it with the data. In the first approximation, the GPD
is proportional to the imaginary part of a DVCS amplitude,
therefore, as discussed in \cite{Magas}, the knowledge (or
experimental reconstruction) of the DVCS amplitude may partly
resolve the problem, provided the phase of the DVCS amplitude is
also known. In other words, a GPD can be viewed as the imaginary
part of an antiquark-nucleon scattering amplitude, or a
quark-nucleon amplitude in the $u$ channel.

Alternatively, one can extract \cite{Burk3, Rad1}, still in a
model-dependent way, the nontrivial interplay between the $x$ and
$t$ dependence of GPD from light-cone wave functions
$$H(x,\xi=0,t)=\int d^2{\bf k}_\bot \psi^*(x,{\bf k}_\bot)\psi(x,{\bf
k}_\bot)+(1-x){\bf q}_\bot),$$ where $\psi(x,{\bf k}_\bot)$ is a
2-particle wave function (see, e.g., \cite{Brodsky}) and $t\equiv
{\bf q}_\bot.$

In two recent papers \cite{GPRV, Kroll} various forms of GPD for
$\xi=0$ were tested against the experimental data on the related
form factors. The agreement with the data in Ref.\cite{GPRV} is
impressive; in Ref.\cite{Kroll} the spatial distribution of
partons in the transverse plane was also calculated. We pursue the
approach of Refs. \cite{GPRV, Kroll} by bringing more arguments
coming from duality in favor of the parametrization for $H(x,t)$
used in \cite{GPRV} and exploring how analyticity and unitarity
affect the $t$ dependence of GPD, the observable form factors and
the calculated distribution of partons in the longitudinal and
transverse planes. The effects are very large. Similarly to papers
\cite{GPRV, Kroll}, we limit ourselves the case of vanishing
skewedness, $\xi=0.$

Regge trajectories play a key role in this analyses. Actually,
there are two groups of trajectories in the problem: one, the
$\rho,$ $ \omega$ etc. trajectories exchanged in the valence quark
distribution function ${\cal H}(x,t)$ (or, equivalently,  the
imaginary part of the $\bar qp$ scattering amplitude). Less
evident are the characteristics of the corresponding trajectory in
"magnetic" densities ${\cal E}(x,t)$ for they cannot be expressed
in terms of any known parton distribution. In Ref. \cite{Kroll}
the $\rho$ trajectory, $\alpha_{\rho}(t)=0.48+0.88t,$ was fitted
to the masses of $\rho(770)$ and $\rho_3(1690)$ and
$\alpha_{\omega}(t)=0.42+0.95t$ to $\omega(782)$ and
$\omega(1690)$. In \cite{GPRV}, instead, the slopes of the
trajectories in ${\cal H}(x,t)$ and ${\cal E}(x,t)$ were fitted to
the data on form factors and are equal to $1.098$ GeV$^2$ and
$1.158$ GeV$^2,$ respectively. The relevant intercepts are
contained in the quark distributions, as discussed in Sec. 5. The
resulting GPDs and related observables are very sensitive to the
above parameters (see Fig. 1 in Ref.\cite{GPRV}). Even more
sensitive are they to any deviation from linear trajectories, as
shown in Sec. 6.

One can start either from trajectories fitted to resonances and
scattering data (parametrizations of non-linear complex mesonic
trajectories fitting the spectra of resonances as well scattering
data exist in the literature \cite{DGMP, Paccanoni}) or treat them
as "effective" ones, to be fitted to the data on form factors. Our
strategy here is to start from trajectories close to those in
\cite{GPRV} (fitted to form factors) and then look for the effects
coming from the deviations from linearity.

The paper is organized as follows. In Sec. 2, following earlier
publications, we show how dual models with Mandelstam analyticity
can be related to deep inelastic scattering and GPD. Regge
trajectories satisfying the analyticity and unitarity constrains
are introduced in Secs. 2 and 3, where their role in the
calculation of form factors is also discussed. The relation
between GPD and form factors is discussed of Sec. 4. A particular
model of GPD with its $t$ dependence determined by dual models
with Mandelstam analyticity, introduced in Sec. 2, is discussed in
Sec. 5. Numerical calculations (form factors and parton
distributions) are presented in Sec. 6. Our (temporary)
conclusions and a discussion can be found in Section 7.

\section{Non-linear Regge
trajectories\\ in dual and composite models} \label{s2}

Dual Amplitudes with Mandelstam analyticity (DAMA) were suggested
(see \cite{DAMA} and earlier references therein) as a way to solve
the manifestly non-unitarity of narrow-resonance dual models
\cite{Veneziano}. The $(u,t)$ term of the crossing-symmetric DAMA
is \eq {D(u,t)=\int_0^1 {dx \biggl({x \over g_1}
\biggr)^{-\alpha(t(1-x))-1}\biggl({1-x \over
g_2}\biggr)^{-\alpha(ux)-1} }, \label{dama}} where $u$ and $t$ are
the Mandelstam variables, and $g_1,\ g_2$ are parameters,
$g_1,g_2>1.$ In what follows we set, for simplicity,
$g_1=g_2=g_0$. Similar expressions are valid for the $(st)$ and
$(su)$ terms. They are not unique since the integrand of
(\ref{dama}) can be multiplied by functions of the type
$f(t(1-x))f(ux).$ Furthermore, the powers in the integrand can be
shifted by integers determined by the quantum numbers of a
particular reaction and relevant exchanges.

The functions $\alpha(y,x)),\ \ y=s,\ t,\ u,$ called in
\cite{DAMA} homotopies, map the physical Regge trajectories
$\alpha(y)$ onto linear functions $a+by$. Contrary to the
narrow-resonance (=linear trajectories) Veneziano amplitude
\cite{Veneziano}, applicable only to soft collisions of extended
objects (strings), and decreasing exponentially at any scattering
angle, DAMA does not only allow for, but even requires the use of
nonlinear Regge trajectories. It will be convenient to write
$\alpha(y)=\alpha(0)+\tilde\alpha(y),$ where $\tilde\alpha(y)$
will denote the nonlinear part of the trajectory.

For $|u|\rightarrow\infty$ and fixed $t,$ DAMA is Regge-behaved
$$D(u,t)\simeq g_0^2(-g_0u)^{\alpha(t)}[G(t)+...],$$
where $$G(t)=\int_0^{\infty}dyy^{-y}y^{-\alpha(t)-1},$$ provided
\cite{DAMA}
$$\left |{\alpha(u)\over{\sqrt{u}\ln u}}\right|\mathop{\leq}\limits_{u\to \infty} const,$$ which is
equivalent to saying that the real part of the trajectory is
bounded \footnote{This basic property of Regge trajectories was
derived \cite{Predazzi}, before the advent of DAMA. For a review
of general properties of Regge trajectories see Ref.
\cite{Trushev}}. Compatibility with the wide-angle scaling
behavior of the amplitude, typical of point-like constituents,
lowers this growth to a logarithm. Examples will be presented
below.

The pole structure of DAMA \eq
{D(u,t)\sim\sum_{l=0}^n{C_{n-l}(t)\over{[n-\alpha(u)]^{l+1}}},
\label{poles}} where $C_{n-l}(t)$ is the residue, whose form is
fixed by the dual amplitude (see \cite{DAMA}), is similar to that
of the Veneziano model except that multiple poles appear on
daughter levels \cite{DAMA}. The pole term (2) in DAMA, comprising
a whole sequence of resonances lying on a complex trajectory
$\alpha(u),$ is a generalization of the Breit-Wigner formula. Such
a "reggeized" Breit-Wigner model has little practical use in the
case of linear trajectories, resulting in an infinite sequence of
poles, but it becomes a powerful tool in case of complex
trajectories with a limited real part and hence a limited number
of resonances.

The threshold behavior of DAMA satisfying the unitarity constrains
\cite{Barut}
$$ D(u,t)\mathop{\sim}\limits_{t\to t_0}\sqrt{t-t_0}[const+\ln(1-t_0/t)],
$$ is correlated with that of the trajectories \cite{DAMA},
$$\Im\alpha(t)\mathop{\sim}\limits_{t\to t_0}(t-t_0)^{\alpha(t_0)+1/2}.$$
For a light threshold this is close to the square-root behaviour
to be used below.

A simple model, compatible both with to the above threshold
behavior and with the Regge asymptotics (or polynomial
boundedness) of the amplitude, yet fitting the data on resonances
spectra, can be made of a sum of square roots \footnote{An
alternative could be: $\alpha(t)\sim-{a+bt\over\sqrt{t_0-t}}
\cite{Bugrij}.$} (the assignment of the signs is uniquely
determined by the requirement of positivity of the imaginary part
etc., see \cite {Trushev}) \cite{Bugr-Kobyl} \eq
{\alpha(t)=\alpha(0) -\sum_i\alpha_i(\sqrt{t_i-t}-\sqrt{t_i}).
 \label{traj}}
 Linear trajectories appear as the
 limiting case of an infinitesimally heavy threshold $t_1\rightarrow\infty$ in
$$ \alpha(t)=\alpha_0-\alpha_1(\sqrt{t_1-t}-\sqrt{t_1})
 $$
 with fixed forward slope of the trajectory
 $\alpha'={\alpha_1\over{2\sqrt{t_i}}}$. The limit of the infinitely rising
 linear trajectory is associated with a hadronic string,
 while the finite value of $t_i$ can be
 interpreted as a barrier where the string breaks (its tension ${1\over
 \alpha'(t)}$ vanishes as $t\rightarrow t_i,$ producing new particles instead of
heavier resonances, see also \cite{burak}).

 The number of thresholds is model-dependent: while the
 lightest one gives the dominant contribution to the imaginary part,
 those heavier promote the rise of the real part, terminating
 at the heaviest threshold.

To illustrate the aforesaid, a toy model can be constructed from
the sum of two thresholds \eq  {\label{Bes2} \alpha(s)=\alpha(0)
-\alpha_1(\sqrt{t_1-t}-\sqrt{t_1})-\alpha_2(\sqrt{t_2-t}-\sqrt{t_2}),}
where $\sqrt{t_1}$ is the lightest one allowed by quantum numbers,
i.e. the two-pion threshold with $t_1=4m_{\pi}^2,$ corresponding
to a loop diagram it the $t$ channel \cite{C-T_Anselm}. The heavy
threshold is chosen phenomenologically: by setting $t_2=4M^2=16
GeV^2,$ we impose an upper bound on the highest mass (slightly
below $2$ GeV) and spin (J=5) resonance lying on the given
trajectory. The parameters $\alpha_1=0.6 GeV^{-1}$ and
$\alpha_2=5.5 GeV^{-1}$ here were chosen such as to match to slope
of the linear trajectory $\alpha(t)=0.5+t,$ see Fig. 1.
\begin{figure}[tbh!]
\centerline{\psfig{figure=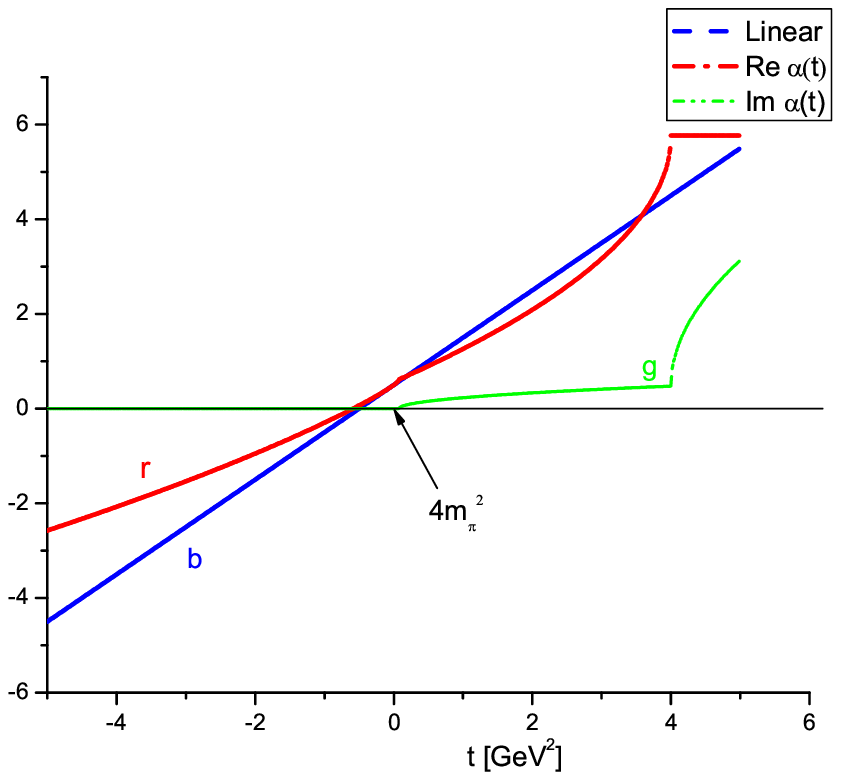,width=0.5\textwidth}
\psfig{figure=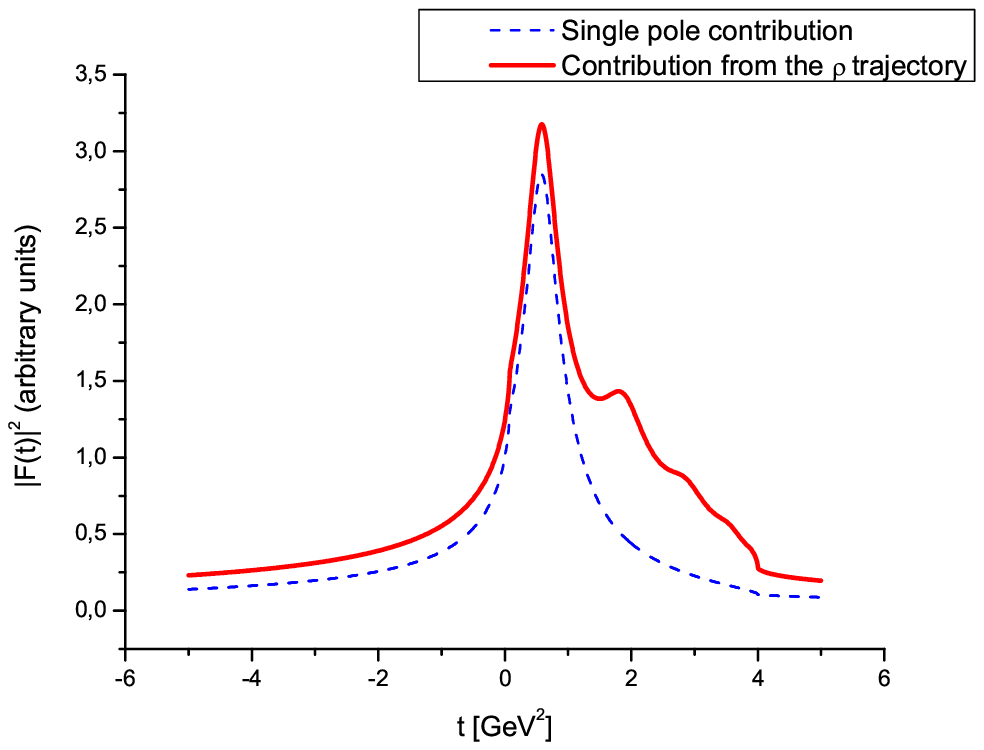,width=0.5\textwidth}} \caption{Left panel:
typical behavior of an analytic trajectory (4) (of its real, read
(r) line, and imaginary, green (g) line, parts) compared with a
linear one, $\alpha(t)=0.5+t.$ Right panel: Form factor modulus
(16) saturated by a single $\rho$ meson pole (blue dotted line)
and the $\rho$ trajectory (4), red solid line. The form of the
latter does not change when summation in (12) is extended to
whatever large $j$, see Sec. 3.} \vspace{0.5cm}
\end{figure}
Notice that $M$ does not correspond to any physical resonance;
rather it is a parameter to be fitted \cite{Bugr-Kobyl} to the
resonances' spectra, as well as to the scattering data
\cite{DGMP}. The construction of Regge trajectories satisfying
theoretical constraints on the threshold- and asymptotic behavior,
yet compatible with the experimental data, is a highly nontrivial
problem. Simple models, like (4), may be helpful as a guide in a
semi-quantitative analysis, as in Sec. 6. In a more rigorous
approach of Ref. \cite{Paccanoni} the real and imaginary parts of
the trajectories were related by a dispersion relation combined
with the unitarity constraints on the threshold behavior and fits
to the resonances' masses and decay widths.

In the limit $|u|, |t|\rightarrow \infty, u/t=const,$ DAMA
(\ref{dama}) scales iff its trajectories have logarithmic
asymptotics. The simplest trajectory that combines the nearly
linear behaviour at small $t$ with a square-root threshold and
logarithmic asymptotics is
\eq{\alpha(t)=\alpha(0)-\gamma\ln\biggl({1+\beta\sqrt{t_0-t}\over{1+\beta\sqrt{t_0}}}\biggr).
\label{log_tr}} More thresholds (introducing more parameters,
however) can be added. Asymptotically,
$\alpha(t)\mathop\simeq\limits_{t\to \infty} -{\gamma\over
2}\ln(-t).$ The asymptotic (scaling) limit of DAMA can be easily
calculated \cite{Chikovani} by the saddle-point method, the saddle
point, for asymptotically logarithmic trajectories, being located
at
$$x_0={\alpha(u)\over{\alpha(u)+\alpha(t)}}=1/2.$$
In this limit, the $(ut)$ term of the amplitude (1) goes like
\cite{Chikovani} \eq{D(u,t)\sim (ut)^{-\gamma\ln(2g)/2}.
\label{scaling}}

The power in (\ref{scaling}) can be fixed by the quark counting
rules \cite{MMT}, by which
$$ln(2g)=2n-1,$$
where n is the number of constituents in a collision. These
numbers should not be taken literally (more details can be found
in \cite{J_NP}) since they may have more relevance to the leading,
vacuum trajectory, while in GPD to be discussed below the main
contribution comes from subleading trajectories.

An interesting link between the fixed scattering scattering angle
regime of DAMA and composite particle models in the infinite
momentum frame \cite{GBB} was established by M.~Schmidt in Ref.
\cite{Schmidt} \footnote{Schmidt \cite{Schmidt} uses trajectories
with a constant asymptotic limit $\ln(-t)\rightarrow const,$
relying on simple arguments of the wide-angle Regge behaviour in
$s^{\alpha(t)\rightarrow const}$ . Less trivial and more relevant
logarithmic trajectories, required by the fixed angle behavior in
DAMA, appeared later \cite{Chikovani} (see also \cite{CGBB}).
Another difference between the ans\"atze used in \cite{Schmidt}
and here is the appearance of the constant $g_0$ (for more details
on the homotopies and the role of $g_0$ see \cite{DAMA}).}.
\begin{figure}[htb]
\center{\includegraphics[width=0.5\textwidth]{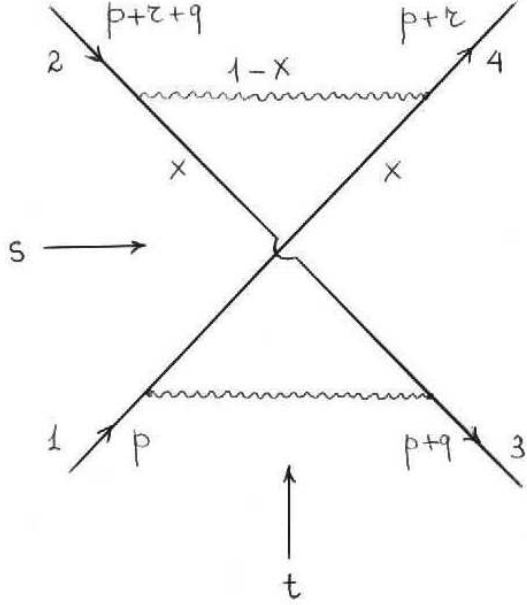}}
\caption{Diagram for (ut) channel elastic scattering,
corresponding to eq. (7)}\vspace{0.5cm}
\end{figure}

In the model of Gunion, Brodsky and Blankenbecler (GBB)
\cite{GBB}, the scattering amplitude corresponding to graph shown
in Fig. 2, is given by \cite{GBB} \eq{
A(u,t)=\int_0^1{dx\over{x^2(1-x)^2}}\int
d^2k_{\bot}\Delta\psi_1(k_{\bot})\psi_2(k_{\bot}+\nonumber\\
(1-x)q_{\bot}-xr_{\bot})\psi_3(k_{\bot}+(1-x)q_{\bot})\psi_4(k_{\bot}+xr_{\bot}),
\label{Schmidt}} where the transverse vectors $r_{\bot}$ and
$q_{\bot}$ satisfy the conditions $r_{\bot}\cdot q_{\bot}=0,\ \
u=-r_{\bot}^2,\ \ t=-q_{\bot}^2,$ in the infinite momentum frame
$$p_1=\biggl(P+{m^2\over{2P}},0,P\biggr),$$
$$p_2=\biggl(P+{{m^2+q^2_{\bot}}\over{2P}},q_{\bot},P\biggr),$$
etc., $P\rightarrow+\infty,$ and $\psi_i$ are quark bound state
wave functions \cite{GBB}.

The form factor $F(q^2)$ and the DIS structure function $F_2(x)$
in terms of the wave functions $\psi$ are \cite{Schmidt} $$
F(q^2)=\int_0^1{dx\over{x(1-x)}}\int
d^2k_{\bot}\psi(k_{\bot})\psi(k_{\bot}+ (1-x)q_{\bot})$$ and
$$F_2(x)=\int d^2k_{\bot}{\psi^2(k_{\bot})\over{1-x}}. $$

     In the limit $s\approx-u=r^2_{\bot}$ and fixed
$t=-q^2_{\bot}$ $$
A(u,t)=\int_0^1{dx\over{x^2(1-x)^2}}g(r^2_{\bot},x)\int
d^2k_{\bot}{{\psi(k_{\bot})\psi(k_{\bot}+
(1-x)q_{\bot})}\over{x(1-x)}} $$ follows\cite{Schmidt}, where
$g(r^2_{\bot},x)\approx\psi(xr_{\bot}),$ while for large $u$ and
fixed $t$ one gets \cite{Schmidt}$$
A(u,t)\sim\int_0^1{dx\over{x(1-x)}}g(r^2_{\bot})
x^{-\alpha(t(1-x))}f(t(1-x)).$$

The parallels between dual (\ref{dama}) and composite models offer
at least three lessons:

1) Apart from "soft" collisions of hadrons (strings)
\cite{Veneziano}, DAMA implicitly contains also the dynamics of
hard scattering of the constituents;

2) the variables appearing in combinations like $t(1-x)$, should
be used in constructing $t$-dependent parton distributions;

3) Regge trajectories are non-linear, complex functions with
well-defined constrains.

\section{Form factors; analyticity and unitarity}

In this section we present general properties of the form factors
with emphasis on their analytic properties and connection with
Regge trajectories, to be utilized in subsequent sections.

There are various choices for the nucleon electromognetic form
factors (ff), such as the Dirac and Pauli ff, $F_1^p(t),\ \
F_1^n(t)$ and $F_2^p(t),\ \ F_2^n(t),$ the Sachs electric and
magnetic ff, $F_E^p(t),\ \ F_E^n(t)$ and $F_M^p(t),\ \ F_M^n(t),$
or isoscalar and isovector electric and magnetic ff, $F_E^s(t),\ \
F_E^v(t)$ and $F_M^s(t),\ \ F_M^v(t)$, where $t=-Q^2$ is the
squared momentum transfer of the virtual photon \cite{Rekalo}.

The Dirac and Pauli form factors are obtained from a decomposition
of the matrix element of the electromagnetic (e.m.) current in
linearly independent covariants made of four-momenta, $\gamma$
matrices and Dirac bispinors as follows
$$<N|J_{\mu}^{e.m.}|N>=e\bar u(p')[\gamma_{\mu}F_1^N(t)+{i\over{2m}}\sigma_{\mu\nu}(p'-
p)_{\nu}F_2^N(t)]u(p),$$ where $m$ is the nucleon mass. Electric
and magnetic ff, on the other hand, are suitable in extracting
them from the experiment:
$$ \begin{array}{l}
\displaystyle {{d\sigma^{lab}(e^-N\rightarrow e^-N)}\over{d\Omega}}= \\
\displaystyle =
{{\alpha_{e.m.}^2\cos^2(\theta/2)}\over{4E^2\sin^4(\theta/2)}}{1\over{1+\bigl({{2E}\over
{m}}\bigr)\sin^2(\theta
/2)}}\Biggl({{G_E^2-{t\over{4m^2}}G^2_M}\over{1-{t\over{4m^2}}}}-2{t\over{4m^2}}G^2_M\tan^2(\theta
/2)\Biggr), \end{array} $$ where $\alpha_{e.m.}=1/137,\ \ E$ is
the incident electron energy, and
$$\sigma_{tot}^{e.m.}(e^+e^-\rightarrow N\bar N)=
{{4\pi\alpha_{e.m.}^2\beta}\over{3t}}[|G_m(t)|^2+{{2m^2}\over
t}|G_E(t)|^2],\ \ \beta=\sqrt{1-{{4m^2}\over t}},$$ or
$$\sigma_{tot}^{e.m.}(p\bar p\rightarrow
e^+e^-)={{4\pi\alpha_{e.m.}^2\over{3p_{c.m.}}\sqrt{t}}}[|G_M(t)|^2+{2m^2\over
t}|G_E(t)|^2],$$ where $p_{c.m.}$ is the proton momentum in the
c.m. system.

The four independent sets of form factors are related by \eq
{G_E^p(t)=G_E^s(t)+G_E^v(t)=F_1^p(t)+\tau^p
F_2^p(t)=[F_1^s(t)+F_1^v(t)]+ \tau^p [F_2^s(t)+F_2^v(t)],} \eq
{G_M^p(t)=G_M^v(t)+G_M^v(t)=F_1^p(t)+F_2^p(t)=[F_1^s(t)+F_1^v(t)]+
[F_2^s(t)+F_2^v(t)],} \eq
{G_E^n(t)=G_E^s(t)-G_E^v(t)=F_1^n(t)+\tau^n
F_2^n(t)=[F_1^s(t)-F_1^v(t)]+ \tau^n[F_2^s(t)-F_2^v(t)],} \eq
{G_M^n(t)=G_M^s(t)-G_M^v(t)=F_1^n(t)+F_2^n(t)=[F_1^s(t)-F_1^v(t)]
+[F_2^s(t)-F_2^v(t)], } where $\tau^{p(n)}={t\over{4m_{p(n)}^2}}.$
They satisfy the normalization conditions  $$ G_E^p(0)=1;\
G_M^p(0)=1+\mu_p;\ G^n_R(0)=0; G^n_M(0)=\mu_n;$$
$$G_E^s(0)=G_E^v(0)={1\over 2};\ G_M^s(0)={1\over
2}(1+\mu_p+\mu_n);\ G_M^v(0)={1\over 2}(1+\mu_p-\mu_n);$$
$$F_1^p(0)=1;\ F_2^p(0)=\mu_p;\ F_1^n(0);\ F_2^n(0)=\mu_n;$$
$$F_1^s(0)=F_1^v(0)={1\over 2};\ F_2^s(0)={1\over 2}(\mu_p+\mu_n);\
F_2^v(0)=(\mu_p-\mu_n),$$ where $\mu_p$ and $\mu_n$ are the proton
and neutron anomalous magnetic moments, respectively.

A basic ingredient of the existing models of form factors
\cite{Rekalo} is the dominance of the $\rho$ meson pole, resulting
e.g. for the isovector meson form factor to the expression
$$G_v(t)={g_{\rho}^2\over{1-t/m_{\rho}^2}},$$ where $g_{\rho}$ is a
constant proportional to the product of $\gamma\rho$ and $\gamma
NN$ couplings. The next step is to include \cite{Dubnicka} other
vector mesons, as well as their excitations, such as  the
isovector: $\rho(770),\ \rho'(1450),\ \rho''(1700)$ and isoscalar
resonances $\omega(782),\ \omega'1420),\ \omega''(1600),\
\phi(1020),\ \phi(1680)$ found in the Review of Particle Physics.

 The use of the trajectories implies a single "dual"
variable instead of the parameters of individual resonances. The
approach of Ref. \cite{Paccanoni} combines the concept of Regge
trajectories with analyticity, unitarity and resonance data
analysis. An economic way to account for the exciting states is to
use Regge trajectories, advocated in the present paper and
bringing us close to dual models. Soon after the discovery of
Veneziano's dual model \cite{Veneziano}, attempts were made
\cite{Frampton} to apply it to form factors. Assuming an infinity
number of neutral vector mesons with the sequence of squared
masses $m^2(n)=m_0^2+nm_1^2,\ n=0,1,...$ the following expression
proton magnetic form factor was derived \cite{Frampton}
$${G_{M_p}(t)\over{\mu_p}}={{\Gamma[1-\alpha(t)]\Gamma[c-\alpha(0)]}
\over{\Gamma[1-\alpha(0)]\Gamma[c-\alpha(t)]}},$$ where
$\alpha(t)$ is the $\rho$ trajectory and $c=3.27.$

Unitarity constrains the threshold behavior of the Regge
trajectories \cite{Barut} (see also \cite{Paccanoni} and
references therein) as well as that of the form factors
\cite{Frazer}
$$\Im F_{\pi}(t)\mathop{\sim}\limits_{t\to t_0}
(t-t_0)^{1-\Re\alpha(t_0)}\Im\alpha(t)\sim(t-t_0)^{3/2},$$ where
$t_0=4\pi^2_{\pi}.$ Any finite width of the resonances requires an
imaginary part to be added, as for example was done in Ref.
\cite{Wataghin}
$${G_{M_p}(t)\over{\mu_p}}=\sum_{i=\rho,\omega,\phi}a_i\Bigl(t-m_i^2+\gamma_i\sqrt{b_i^2-t}\Bigr)^{-1},$$
where $\gamma_i$ are resonances' widths and the fitted values of
the parameters $b_i$ are: $b_{\rho}=0.28$ GeV, $b_{\omega}=0.42$
GeV, $b_{\phi}=0.99$ GeV. This approach is close in spirit to that
based on dual analytic model, introduced in the previous section,
and illustrated in Figs. 1, showing the calculated modulus of the
form factor resulting from a single $\rho$ pole contribution
(dotted line) and from a sequence of poles generated by trajectory
(4) \eq{{G_{M_p}\over{\mu_p}}=\left\vert\sum_{j=1}^{\infty}
{(0.5)^j\over{(j - \alpha(\rho))}}\right\vert. \label{FF}} This
sum is similr to the pole decomposition of the dual amplitude,
namely it is a sum of "reggeized" {implied by duality)
Breit-Wigner resonances (cf. (2)). The upper limit of summation
includes the highest resonance lying on the $\rho$ trajectory (4),
however the large-$|t|$ behaviour of the form factor, Fig. 3, is
not affected by higher spin values (here, $j>5$, up to infinity),
from where the real part of the trajectory does not contribute any
more. The inclusion of a large (infinite) number of poles
appearing on the second sheet is important in dual models
\cite{DAMA}. The non-appearance of higher resonances and the
transition to a smooth continuum can result either from an upper
bound on the real part of the trajectory, as is  DAMA \cite{DAMA},
or from the rapid rise of its imaginary part, making the
resonances unobservable.

Similarly to the Veneziano model, form factors were derived
\cite{Prognimak} from DAMA (1): \eq{F(t)_{\pi}=\int_0^1 dx
x^{-\alpha(t(1-x))}(1-x)^{-1+n}, \label{Progn}} where $n$ is an
integer providing the correct (according to quark counting rules)
large-$|t|$ behavior of the form factor.

Calculations of form factors from GPD, Sec. 4-6,  are more
involved and less predictable, especially in the case of complex
Regge trajectories.

\section{ Generalized parton distributions and form factors}
\label{FormF}

Form factors are related to generalized parton distributions (GPD)
by the standard sum rules \cite{Rad, Ji}
$$F_1^q(t)=\int_{-1}^1 dx H^q(x,\xi,t),$$
$$F_1^q(t)=\int_{-1}^1 dx H^q(x,\xi,t).$$

The integration region can be reduced to positive values of
$x,~0<x<1$ by the following combination of non-forward parton
densities \cite{Rad1,GPRV} $${\cal
H}^q(x,t)=H^q(x,0,t)+H^q(-x,0,t),$$ $${\cal
E}^q(x,t)=E^q(x,0,t)+E^q(-x,0,t),$$ providing
\eq{F^q_1(t)=\int_0^1 dx {\cal H}^q(x,t),\label{01}}
\eq{F^q_2(t)=\int_0^1 dx {\cal E}^q(x,t).\label{02}}

The proton and neutron Dirac form factor are defined as $$
F_1^p(t)=e_uF_1^u(t)+e_dF_1^d(t),$$ $$
F_1^n(t)=e_uF_1^d(t)+e_dF_1^u(t),$$  where $e_u=2/3$ and
$e_d=-1/3$ are the relevant quark electric charges.

In the limit $t\rightarrow 0$ the functions $H^q(x,t)$ reduce to
usual valon quark densities in the proton: $$ {\cal\
H}^u(x,t=0)=u_v(x),\ \ \ {\cal H}^d(x,t=0)=d_v(x)$$ with the
integrals $$\int_0^1 u_v(x)dx=2,\ \ \ \int_0^1 d_v(x)dx=1 $$
normalized to the number of $u$ and $d$ valence quarks in the
proton.

Contrary to ${\cal H},$ the "magnetic" densities ${\cal
E}^q(x,t=0)\equiv {\cal E}^q(x)$ cannot be directly expressed in
term of the known parton distributions, however their
normalization integrals $$\int_0^1{\cal E}^q (x)dx\equiv k_q $$
are constrained by the requirement that the values $F_2^p(t=0)$
and $F_2^n(t=0)$ are equal to the magnetic moments of the proton
and neutron, whence $k_u=2k_p+k_n\approx 1.673$ and
$k_u=k_p+2k_n\approx -2.033$ follows \cite{GPRV}. Explicit
parameterizations for the forward and non-forward structure
functions as will be presented in the next section.

The Fourier-Bessel integral \beq q(x,b)={1\over 2\pi}\int_0^\infty
d\sqrt{-t}J_0(b\sqrt{-t}){\cal H}(x,t) \end{equation} provides a
mixed representation of longitudinal momentum and transverse
position in the infinite-momentum frame \cite{Burk1, Burk2}.

\newpage
\section{Modelling non-forward parton distributions;\\
connection with Regge-dual models}\label{GPD}

The simplest model for the proton non-forward parton density is a
factorized form \beq {\cal H}(x,t)=q_v(x)F_1(t),\eeq where
$q_v(x)$ is the parton density and $F_1(t)$ is the proton form
factor. It trivially reproduces $F_1^p(t)$ and $q_v(x)$ in the
forward limit, but it conflicts both with Regge (R) behavior \beq
{\cal H}_R(x,t)\sim x^{-\alpha(t)}, \eeq{6} valid at small $x,$
and with the light-cone formalism, suggesting a Gaussian (G)
parametrization for non-forward parton densities \cite{BGNPZ,
Rad1} $${\cal H}_G^q(x,t)=q_v(x)e^{-(1-x)t/4x\lambda^2},$$ where
the scale $\lambda^2$ characterizes the average transverse
momentum of the valence quarks in the nucleon. To satisfy the
Drell-Yan-West (DYW) relation \cite{DY, West} between the $x\to 1$
behavior of the structure functions and the $t-$dependence of the
elastic form factors the above expression should be modified e.g.
as \cite{GPRV} \eq{{\cal H}^q(x,t)=q_v(x)x^{-\alpha'(1-x)t},
\label{a}} where $\alpha'$ is the slope of a Regge trajectory
(other modifications of the parton distributions as well as their
 relation to the light-cone wave function of a composite system
 are discussed in \cite{Burk1, Burk2}). Noticing the similarity
between this expression and the relevant factor in the integrand
of (1), we suggest the following parametrization for the
$t-$dependent GPD: \eq{{\cal
H}_G^q(x,t)=q_v(x)(x/g_0)^{-\tilde\alpha(t)(1-x)t}=(x/g_0)^{-\alpha(t)(1-x)}f(x),
\label{b}} where $\tilde\alpha(t)$ is the $t-$dependent part of
the Regge trajectory, $g_0>1$ ($x_0$ in Ref. \cite{Kroll}) is a
parameter defined in Sec. 2 and $f(x)$ is the large-$x$ factor of
the parton distribution (see below). For linear trajectories and
$g_0=1,$ eq. (20) reduces to (19).

Regge behavior $\sim x^{-\alpha(0)}$ of DIS structure functions
and relevant parton distributions at small $x$ is well established
for small and moderate virtualities $Q^2;$ at higher $Q^2$ it is
replaced by QCD evolution (see e.g. \cite{DJP} and references
therein). The values of the "Regge-intercepts" in the parton
distributions may depend on the flavor of the relevant quark. For
example, in the global fits of MRST2002 \cite{MRST}:
\eq{u_v(x)=0.262x^{-0.69}(1-x)^{3.50}(1+3.83x^{0.5}+37.65x),\label{PD1}}
\eq{d_v(x)=0.061x^{-0.65}(1-x)^{4.03}(1+49.65x^{0.5}+8.65x),\label{PD2}}
implying $\alpha(0)=0.69$ in the $u$-quark distribution and $0.65$
in the $d$-quark distribution, which means slightly different
trajectories exchanged in the $t$-channel of the (fictive) $\bar
up$ and $\bar dp$ scattering amplitude, once the GPD is associated
with the imaginary part of the $u-$ channel quark-proton
scattering amplitude. We use these expressions in our calculations
below.

The "magnetic" densities ${\cal E}^q(x,t)$  enter in $F_2(t)$ and
contain new information about the nucleon structure, however they
cannot be directly expressed in terms of any known parton
distribution. Following \cite{GPRV}, we write them in the form
${\cal E}^q(x,t)={\cal E}^q(x)x^{-(1-x)\alpha_E't},\
\alpha_E'=1.158 GeV^{-2},$ similar to ${\cal H}^q(x,t)$ but with
an extra large-$x$ factor, i.e. \eq{{\cal E}^u(x)={k_u\over
{N_u}}(1-x)^{\eta_u}u_v(x)\label{E1}} and \eq{{\cal
E}^d(x)={k_d\over {N_d}}(1-x)^{\eta_d}d_v(x),\label{E2}} with
$\eta_u=1.52$ and $\eta_d=0.31$ fitted \cite{GPRV} to the data.
The remaining constants are fixed by normalization:
$$N_u=\int_0^1 dx (1-x)^{1.52}u(x)=1.53;$$ $$ N_d=\int_0^1 dx
(1-x)^{0.31}d(x)=0.82,$$ whence $k_u\approx 1.673,\ k_d\approx
-2.033.$

\section{Numerical estimates}

Below we illustrate how the nonlinearity (complexity) of the
trajectories and the introduction of $g_0>1$ affect the behavior
of the calculated form factors and quark distributions as
functions of the impact parameter $b$ and of the Bjorken variable
$x.$
\begin{figure}[tbh!]
\centerline{\psfig{figure=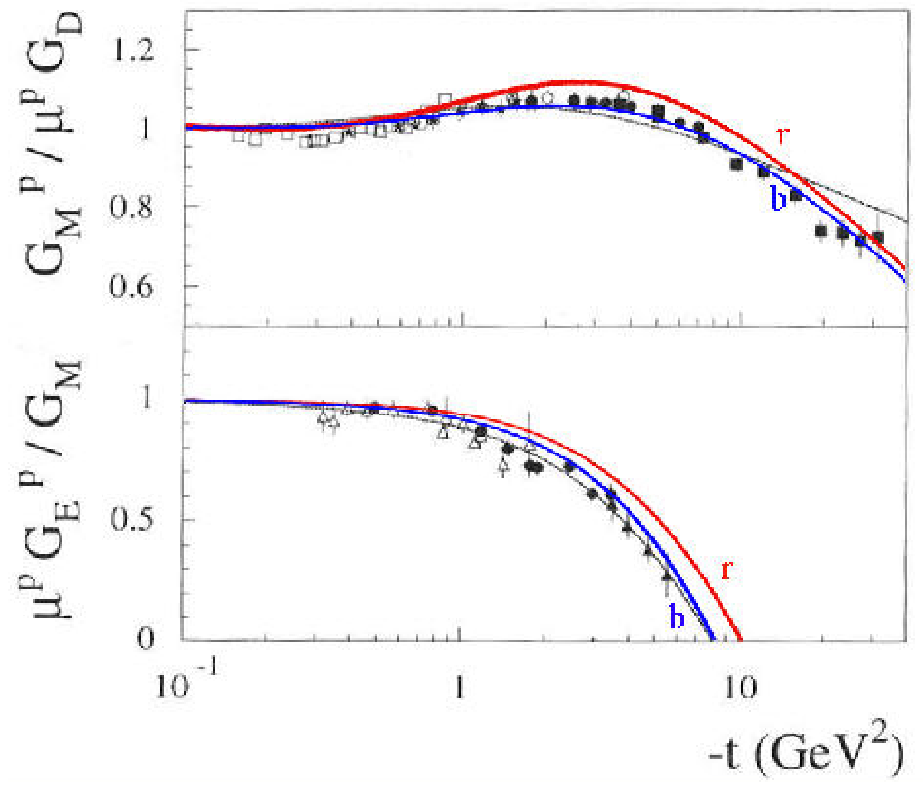,width=0.5\textwidth}
\psfig{figure=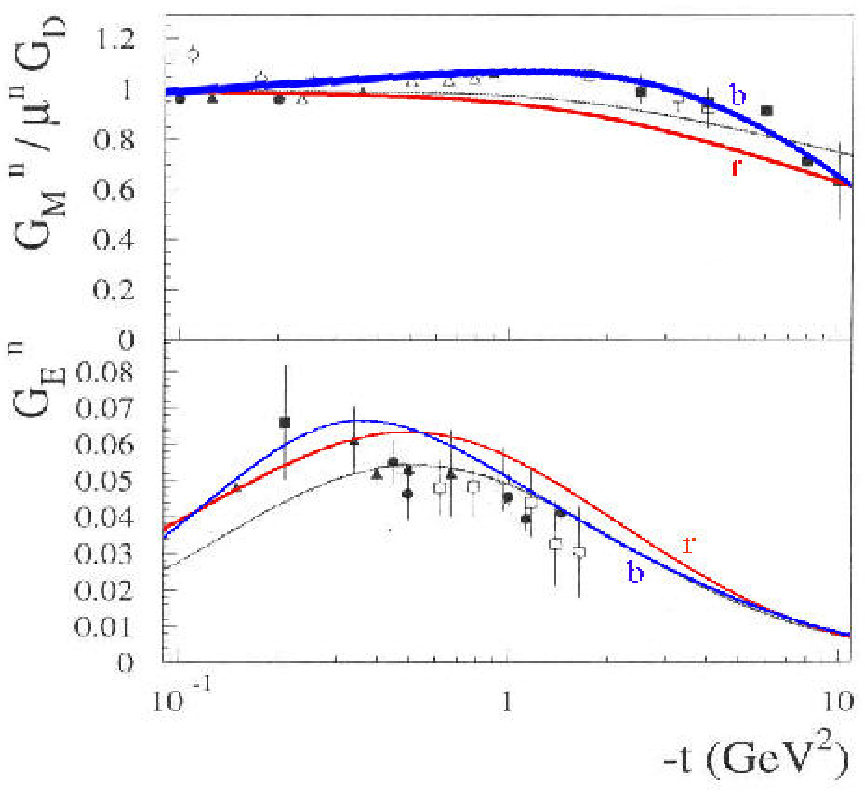,width=0.5\textwidth}}\caption{{\it Left
panel}: Proton magnetic form factor relative to the dipole form
factor (left upper panel) and the ratio of the magnetic to
electric form factors (left lower panel). The black thin curves
correspond to the linear trajectory of Ref. \cite{GPRV}, the red
(r) curve to the nonlinear trajectory (24) with $g_0=1,$ and the
blue (b) curve to the same nonlinear trajectory but with $g=1.05.$
The data for the proton magnetic form factor $G_M^p$ are from
\cite{a} (open squares), \cite{b} (open circles), \cite{c} (solid
stars), \cite{d} (open stars), \cite{e} (solid circles), \cite{f}
(solid squares), and those for the ratio $G^p_E/G^p_M$ are from
\cite{g} (solid circles), \cite{h} (open triangles) and \cite{i}
(solid triangles). {\it Right panel}: Ratio of the neutron
magnetic form factor to the dipole form factor (right upper panel)
and neutron electric form factor (right lower panel) with curve
conventions as on the left panel. The data for the neutron
magnetic form factor $G_M^n$ are from \cite{j} (open circles),
\cite{k} (solid circles), \cite{l} (open triangles), \cite{m}
(solid triangles), \cite{n} (solid squares) and \cite{o} (solid
squares). The data for the neutron electric form factor are from
MAMI \cite{p} (triangles), NIKHEF \cite{r} (solid squares) and
JLab \cite{s} (solid circles) and \cite{t} (open squares).}
\vspace{0.1cm}
\end{figure}

We first explore the effects coming from non-linear Regge
trajectories by calculating several observable form factors and
their ratios using the expressions for the $t-$dependent GPD, eq.
(20). As a reference frame, we use fits from the paper \cite{GPRV}
nicely reproducing the data. In that paper linear trajectories
with the slopes $\alpha_1'=1.098$ GeV$^2$ in ${\cal H}(x,t)$ and
$\alpha_2'=1.158$ GeV$^2$ in ${\cal E}(x,t)$ were used. The
relevant intercepts come from the parton distributions (21), (22),
and they are equal to $0.69$ and $0.65$ in the $u$ quark and $d$
quark distributions, respectively \cite{MRST}. With these
parameters (and $g_0=0,$ matching the model of ref. \cite{GPRV})
one reproduces the results of Ref. \cite{GPRV}, some of them shown
in Fig.3 in thin black line.
\begin{figure}[tbh!]
\centerline{\psfig{figure=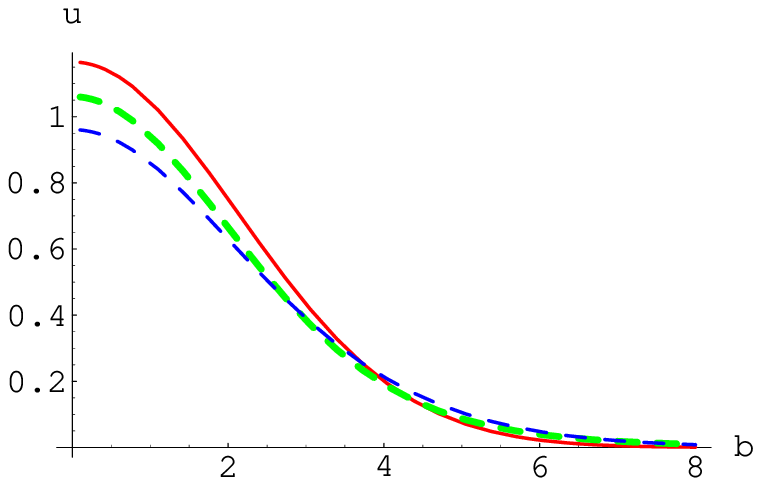,width=0.33\textwidth}
\psfig{figure=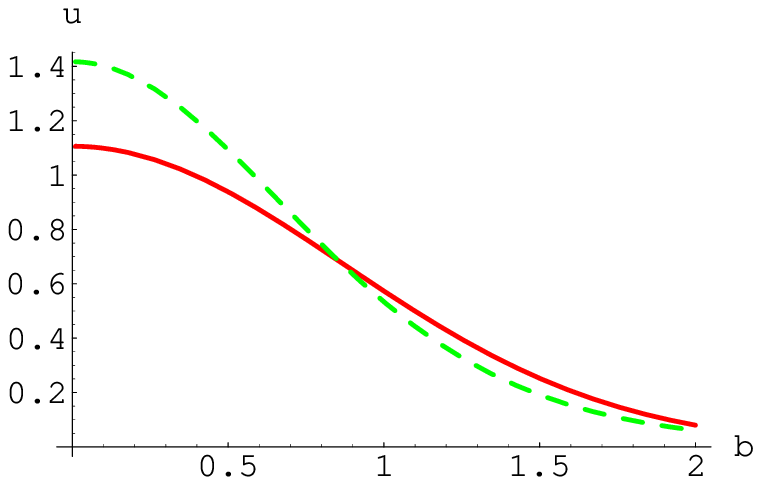,width=0.33\textwidth}\psfig{figure=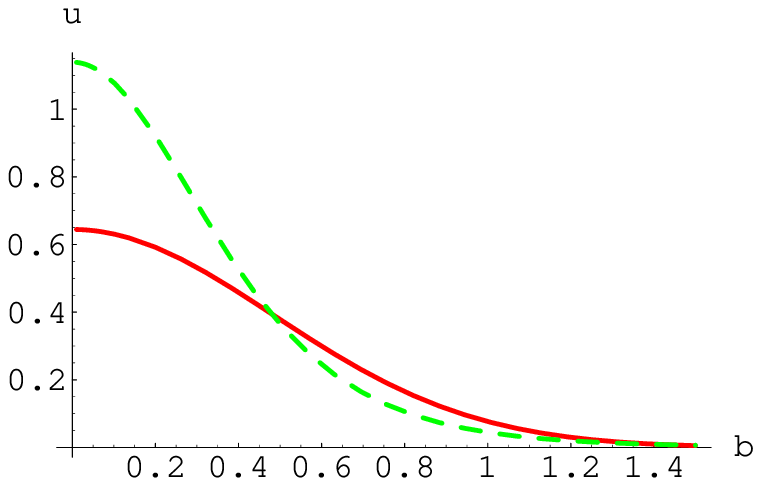,width=0.33\textwidth}}
\caption{Impact-parameter distribution of the $u$ quark matter at
$x=0.1$ (left panel), $x=0.5$ (middle) and $x=0.7$ (right panel).
The curves on the left panel correspond (from the top) to
calculations with: $\tilde\alpha(t)=1.098t$ (red, solid line); eq.
(4) (green, dotted line) and the logarithmic trajectory eq. (5)
with $\gamma=0.3,\ \beta=0.366$ and $t_0=4m_{\pi}^2$ (blue, dotted
line). Curves on the middle and right panel correspond to
calculations with eq. (4) (green dotted line) and the above linear
trajectory (red solid line). Those with the logarithmic trajectory
(5) for $x>0.3$ go off the common trend.} \vspace{0.5cm}
\end{figure}
Next we explore the effects coming from the nonlinearity of the
Regge trajectories, as well as from $g_0>1,$ both introduced in
Sec.2.  By keeping the less known ${\cal E}^q(x,t)$ intact we vary
the trajectory in ${\cal H}^u(x,t)$ and the value of $g_0.$ The
effect proves to be very large. For example, even a minor (with
respect e.g. to (4)) deviation from the linear trajectory,
\eq{\tilde\alpha(t)=1.026t-0.02(\sqrt{4m_{\pi}^2-t}-2m_{\pi}),\label{Nonl}}
with the forward slope $\alpha'\approx 1.098 GeV^{-2}$ matching
the linear trajectory fitted in \cite{GPRV} to the data, gives a
sizable effect, augmenting e.g. the ratio $G_M^p/\mu^pG_D$, as
seen in Fig.3 (red line).
\begin{figure}[tbh!]
\centerline{\psfig{figure=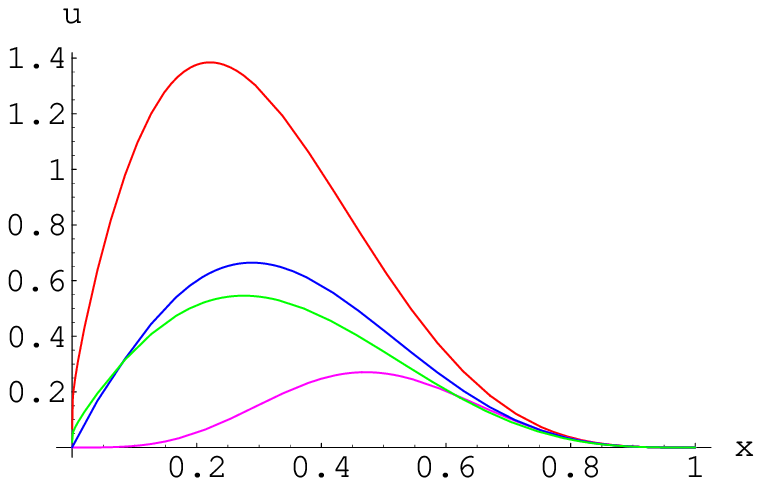,width=0.5\textwidth}
\psfig{figure=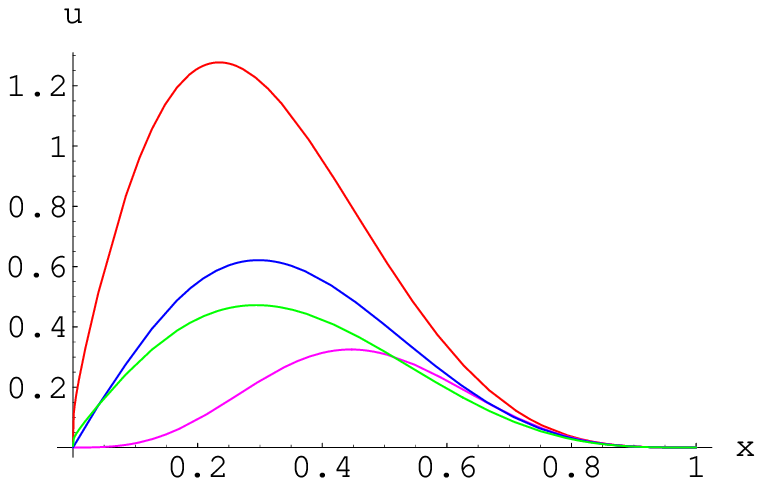,width=0.5\textwidth}} \caption{Longitudinal
momentum distribution of the $u$ quark matter calculated with a
linear trajectory $ \tilde\alpha(t)=1.098t$ (left panel) and eq.
(4) (right panel). The curves (from the top) correspond to:
[$t=-0.7 GeV^2, g_0=1$] (red line), $[t=-0.9 GeV^2, g_0=2]$ (blue
line), $[t=-0.7 GeV^2, g_0=5]$ (green line),  $[t=-3 GeV^2,
g_0=1]$ (pink line).} \vspace{0.5cm}
\end{figure}
The effect from $g_0>1,$ combined with the nonlinear trajectory,
is shown in Fig. 3 (blue line). In general, it compensates the
rise due to the nonlinearity of the trajectory, although the
interplay of these two effects is much more complicated. The use
of analytic trajectories and/or $g_0>1$ in ${\cal E}^q(x,t)$ will
make the situation much more complicated but, but at the same
time, interesting. The use of a "realistic" trajectory like eq.
(4) changes the behavior of the observables dramatically,
requiring a complete rearrangement of the model or, at least, of
its parameters.

Fig. 4 shows the $u$ quark impact parameter distribution
calculated from eqs. (16), (20) at three fixed values of $x=0.1,\
0.5$ and $0.7$ and for three representative trajectories: linear,
$\tilde\alpha(t)=1.098t$ (red (r) curves), "square root", eq. (4)
(green (g) curves) and logarithmic, eq. (5) with $\gamma=3,\
\beta=0366 GeV^{-2}$ and $t_0=4m_{\pi}^2$ (blue (b) curves). The
calculated distributions depend dramatically on the form of the
trajectories, especially near the endpoints $x=0$ and $1$.

The role of the parameter $g_0,$ combined with the variation of
the trajectories, can be seen also in Fig. 5, where the $u$ quark
distribution is plotted against $x$ for several fixed values of
$t$ and of $g_0$ and two typical trajectories: a linear one,
$\tilde\alpha(t)=1.098t,$ and (\ref{Bes2}). Here the dependence on
the form of the trajectories is less pronounced then in Fig. 4,
where integration in $t$ is involved.

Given the available freedom in the choice of the trajectories, the
present calculations can serve only as an indication of the
existing trends, rather then fits or predictions to the data.

\section{Conclusions and outlook}

With this paper we wish to emphasize the important role of the
analytic properties of the strong interaction theory, manifest
here in the form of the Regge trajectories. The wide-spread
prejudice that the trajectories are linear has different sources:
1) the masses of the resonances lie on approximately linear
trajectories; 2) the Veneziano and string models \cite{Veneziano}
provide a theoretical basis in favor of this behavior; 3) relevant
calculations are simple. On the other hand, the theory demands
that the trajectories be analytic functions of their arguments
with threshold singularities imposed by unitarity, and that the
asymptotic behavior be compatible with the polynomial boundedness
of the amplitude. Trajectories satisfying these constraints and
fitting the data both for positive (particles spectra) and
negative (scattering data) values of their arguments are known
from the literature (see, e.g., \cite{DGMP, Paccanoni}). In this
paper we show that they affect considerably the calculated GPD and
their moments. More work is needed to specify and quantify the
role of various (parent and daughter) subleading  trajectories (of
poles and cuts).

Another message of this paper is that, in a certain kinematical
region, the integrand of the dual amplitude with Mandelstam
analyticity (DAMA) (1) can be identified with a generalized parton
distribution (GPD), the integration variable being associated with
the parton longitudinal momentum, as suggested in \cite{Protvino}.
For fixed $Q^2$ and $s,$ the integrand of DAMA (1) (a GPD?) has
the form
\eq{(x/g_0)^{-\tilde\alpha(t(1-x))-n}((1-x)/g_0)^m,\label{Concl}}
where $g_0>1,$ and $n$ and $m$ are reaction-dependent constants.
The first factor in (26) is the small-$x$ Regge-behaved term,
while the second one is the familiar large-$x$ term. An immediate
observation is the similarity between the moments of the GPD (form
factors), eq. (14) and expression (13), derived from DAMA. I
wonder if this could mean a bootstrap relation between the
antiquark-hadron scattering amplitude (GPD) and the hadron-hadron
amplitude resulting from integration (1) of a GPD ? The appearance
of $x$ in the exponents of the (modified) structure functions, (or
parton distributions) may have interesting consequences by itself
-- both for theory and phenomenology. In any case, a better
understanding of the physical meaning of the variables appearing
in GPD, especially with non-zero skewedness, $\xi\neq 0,$ is
needed.

The approach in this paper combines elements of the analytic $S-$
matrix theory, namely Regge poles and duality, known to be
efficient for "soft" collisions at large distances, with the
small-distance partonic picture. The interface and merge of these
seemingly orthogonal approaches may bring new ideas about the
transition form perturbative to non-perturbative physics. Much of
this information is encoded in the form of the complex Regge
trajectories. Recently, explicit models for deeply virtual Compton
scattering amplitudes (DVCS) appeared in the literature
\cite{DVCS}. Their imaginary part can provide additional
information about GPD.

In a perspective one can think of extending the asymptotic Regge
pole model to the low-energy resonance region by incorporating a
dual amplitude, e.g., DAMA. The crossing-symmetric properties of
dual amplitudes will make possible the inclusion of the $t>0$
region in DVCS and resulting GPD. It may also help to connect the
high-energy (Regge) region with the low-energy (resonance) domain,
where new data from JLab is expected.

\section*{Acknowledgements}

 I am grateful to A.I. Bugrij, V.K. Magas and F. Paccanoni for
useful discussions on analyticity and partonic distributions.

\vfill \eject

\end{document}